\documentclass[twocolumn,english,aps,prl,superscriptaddress]{revtex4-1}
\usepackage[latin9]{inputenc}
\usepackage{color}
\usepackage{babel}
\usepackage{graphicx}
\usepackage[unicode=true,
 bookmarks=false,
 breaklinks=false,pdfborder={0 0 1},backref=section,colorlinks=false]
 {hyperref}
\usepackage{breakurl}

\makeatletter

\@ifundefined{textcolor}{}
{%
 \definecolor{BLACK}{gray}{0}
 \definecolor{WHITE}{gray}{1}
 \definecolor{RED}{rgb}{1,0,0}
 \definecolor{GREEN}{rgb}{0,1,0}
 \definecolor{BLUE}{rgb}{0,0,1}
 \definecolor{CYAN}{cmyk}{1,0,0,0}
 \definecolor{MAGENTA}{cmyk}{0,1,0,0}
 \definecolor{YELLOW}{cmyk}{0,0,1,0}
}

\usepackage{babel}
\usepackage{lineno}
\usepackage{epstopdf}
\usepackage{epsfig}
\usepackage{epstopdf}
\usepackage{bmpsize}
\usepackage{amsfonts}

\newcommand{\ben}{\begin{eqnarray}}
\newcommand{\een}{\end{eqnarray}}

\newcommand{\bef}{\begin{figure}[h!bt]\centering}
\newcommand{\eef}{\end{figure}}
\newcommand{\bet}{\begin{table}[hbt]\centering}
\newcommand{\eet}{\end{table}}

\makeatother

\begin{document}

\title{Coexistence of superconductivity and antiferromagnetism in Ca$_{0.74(1)}$La$_{0.26(1)}$(Fe$_{1-x}$Co$_{x}$)As$_{2}$
single crystals}

\author{Shan Jiang}

\affiliation{Department of Physics and Astronomy and California NanoSystems Institute,
University of California, Los Angeles, CA 90095, USA}

\author{Lian Liu}

\affiliation{Department of Physics, Columbia University, New York, NY 10027, USA}

\author{Michael Sch\"utt }

\affiliation{School of Physics and Astronomy, University of Minnesota, Minneapolis,
MN 55455, USA}

\author{Alannah M. Hallas}

\affiliation{Department of Physics, McMaster University, Hamilton, Ontario, L8S
4M1, Canada}

\author{Bing Shen}

\affiliation{Department of Physics and Astronomy and California NanoSystems Institute,
University of California, Los Angeles, CA 90095, USA}

\author{Wei Tian}

\affiliation{Quantum Condensed Matter Division, Oak Ridge National Laboratory,
Oak Ridge, TK 37831, USA}

\author{Eve Emmanouilidou}

\affiliation{Department of Physics and Astronomy, University of California, Los
Angeles, CA 90095, USA}

\author{Aoshuang Shi}

\affiliation{Department of Physics and Astronomy, University of California, Los
Angeles, CA 90095, USA}

\author{Graeme M. Luke}

\affiliation{Department of Physics, McMaster University, Hamilton, Ontario, L8S
4M1, Canada}

\author{Yasutomo J. Uemura}

\affiliation{Department of Physics, Columbia University, New York, NY 10027, USA}

\author{Rafael. M. Fernandes }

\affiliation{School of Physics and Astronomy, University of Minnesota, Minneapolis,
MN 55455, USA}

\author{Ni Ni}

\email{Corresponding author: nini@physics.ucla.edu}

\affiliation{Department of Physics and Astronomy and California NanoSystems Institute,
University of California, Los Angeles, CA 90095, USA}
\begin{abstract}
We report the transport, thermodynamic, $\mu$SR and neutron study
of the Ca$_{0.74(1)}$La$_{0.26(1)}$(Fe$_{1-x}$Co$_{x}$)As$_{2}$
single crystals, mapping out the temperature-doping level phase diagram.
Upon Co substitution on the Fe site, the structural/magnetic phase
transitions in this 112 compound are suppressed and superconductivity
up to 20 K occurs. Our measurements of the superconducting and magnetic
volume fractions show that these two phases coexist microscopically
in the underdoped region, in contrast to the related 10-3-8 Ca$_{10}$(Pt$_{3}$As$_{8}$)((Fe$_{1-x}$Pt$_x$)$_{2}$As$_{2}$)$_{5}$
compound, where coexistence is absent. Supported by model calculations,
we discuss the differences in the phase diagrams of the 112 and 10-3-8
compounds in terms of the FeAs interlayer coupling, whose strength
is affected by the character of the spacer layer, which is metallic
in the 112 and insulating in the 10-3-8.
\end{abstract}
\maketitle
Since the observation of 26 K superconductivity (SC) in LaFeAsO$_{1-x}$F$_{x}$
\cite{1111}, several families of Fe-based superconductors (FBS) have
been discovered. Among them, Ca$_{1-x}$La$_{x}$FeAs$_{2}$ (CaLa112)
with $T_{c}$ up to 42 K, crystalizes in a monoclinic lattice \cite{japan112b,yakita}.
This crystal structure based on the FeAs-(Ca/La)-As-(Ca/La)-FeAs stacking
contains the prototypical FeAs layers made of the edge-sharing FeAs$_{4}$
tetrahedra, as well as As layers made of zig-zag chains. The presence
of these As chains has made CaLa112 unique in many aspects. Our recent
study shows the Ca$_{0.73}$La$_{0.27}$FeAs$_{2}$ compound, which
has an effectively electron overdoped FeAs layer, is the ``parent''
compound of the CaLa112 FBS, highlighting the dual nature of itinerant
and localized magnetism in FBS \cite{shan}. In this material, a monoclinic
to triclinic structural phase transition occurs at 58 K and a paramagnetic
to stripe antiferromagnetic (AFM) phase transition appears at 54 K.
Furthermore, metallic spacer layers are observed via ARPES measurement
\cite{ARPESli,shan}. Besides Ca doping, which adds hole-like carriers,
Co substitution on the Fe sites, which adds electron-like carriers,
can also stabilize SC in Ca$_{0.73}$La$_{0.27}$FeAs$_{2}$ \cite{japan1,japan2,shan,shi1}.
Therefore, as a FBS series with metallic spacer layers and no $C_{4}$
rotational symmetry, the characterization of the Co doped CaLa112
is of particular interest. Here we report a systematic study of Co
doped Ca$_{0.73}$La$_{0.27}$FeAs$_{2}$ (Co-CaLa112). We show that upon Co doping, the structure/magnetic
phase transitions in Ca$_{0.73}$La$_{0.27}$FeAs$_{2}$ are suppressed
and bulk SC up to 20 K emerges. Using the $x$=0.046 sample as a representative,
we present the superconducting properties of the Co-CaLa112 FBS. In
particular, microscopic coexistence of AFM and SC in this system is
revealed by combined $\mu$SR, susceptibility and neutron scattering
measurements. We contrast this behavior with the related 10-3-8 compound
Ca$_{10}$(Pt$_{3}$As$_{8}$)((Fe$_{1-x}$Pt$_x$)$_{2}$As$_{2}$)$_{5}$,
which has similar FeAs interlayer spacing as Co-CaLa112, but displays
no AFM-SC coexistence. We interpret this difference in terms of the
nature of the spacer layer, which is metallic in the 112 compound
but insulating in the 10-3-8 system.

Single crystals of Ca$_{0.74(1)}$La$_{0.26(1)}$(Fe$_{1-x}$Co$_{x}$)As$_{2}$
were grown out of self-flux at the ratio of CaAs : LaAs : FeAs : CoAs
: As = 1.3 : 0.5 : 1-$x$ : $x$ : 0.7 \cite{shan}. Zero field (ZF)
and longitudinal field (LF) $\mu$SR experiments on $\sim$ 200 mg
of single crystals with random orientations were performed in a conventional
Helium gas flow cryostat in TRIUMF, Canada. Single crystal neutron
diffractions were collected on the $x=0.025$ sample using HB-1A triple-axis
spectrometer located at the High Flux Isotope Reactor of the Oak Ridge
National Laboratory.

\begin{figure}
\centering \includegraphics[width=3.4in]{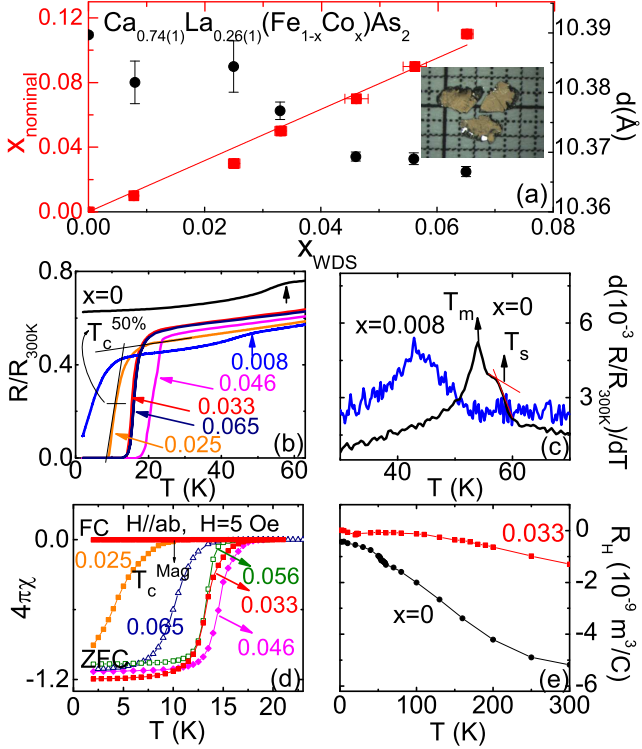} \protect\protect\caption{Ca$_{0.74(1)}$La$_{0.26(1)}$(Fe$_{1-x}$Co$_{x}$)As$_{2}$: (a)
The $x_{\mathrm{nominal}}$ vs. $x_{\mathrm{WDS}}$ and the evolution
of the FeAs interlayer distance $d$ with $x_{\mathrm{WDS}}$.
Inset: single crystals against the 1 mm scale. (b) The temperature
dependent normalized resistance $R/R_{300K}$ for representative samples.
The 50\% criterion to infer $T_{c}$ from resistivity is depicted
for $x=0.025$ sample. (c) The derivative of $R/R_{300K}$ vs. T for
the $x=0$ and $x=0.008$ samples. The criteria to infer $T_{s}$
and $T_{m}$ are depicted. (d) The temperature dependent ZFC and FC
data with $H\parallel ab$. For the $x=0.025$ sample, the value of
$4\pi\chi$ is averaged on four pieces. The criterion to infer $T_{c}$
from susceptibility is depicted. (e) The temperature dependent Hall
coefficient for the $x=0$ and $x=0.033$ samples. }

\label{fig:Fig1}
\end{figure}

Since both La and Co concentrations can vary, wave-length dispersive
spectroscopy (WDS) measurements were performed on at least 5 pieces
in each batch to determine the concentration of the sample. The result
is summarized in Fig. 1 (a), showing that the ratio of $x_{\mathrm{WDS}}$/$x_{\mathrm{nominal}}$
is $\sim$ 0.63. Throughout the paper, we take $x$ to be the $x_{\mathrm{WDS}}$
value. The maximum Co doping achieved is at $x=0.065$ while the growth
with higher Co concentrations is unsuccessful. The concentration of
La in each batch has a random variation as 0.270(4), 0.250(10), 0.269(3),
0.259(5), 0.249(7), 0.264(2) and 0.270(6) for $x=0$, 0.008, 0.025,
0.033, 0.046, 0.056 and 0.065, respectively. Since La concentration
shows some variation, we label it as 0.26(1) in the chemical formula
for simplicity. In each batch, the FeAs interlayer distance is inferred
by measuring the (0, 0, $l$) X-ray peaks diffracted from the $ab$
planes of several single crystals. For $x=0$, the interlayer distance
is $d\approx$10.36 $\AA$, comparable to the value in the 10-3-8
FBS \cite{ni1038}. Upon Co doping, the interlayer distance shrinks,
decreasing by $0.22\%$ at $x=0.065$. This is comparable to the
decrease of the interlayer distance at similar Co doping in BaFe$_{2}$As$_{2}$
\cite{nico}.

Figure 1 (b) shows the representative temperature dependence of $R/R_{300K}$,
highlighting the resistive anomalies in the $x=0$ and $x=0.008$
samples. The derivatives of the $R/R_{300K}$ with temperature for
these two samples are shown in Fig. 1(c). Two-kink features are observed
in d$R/R_{300K}$/dT with the higher temperature kink associated with
the monoclinic to triclinic phase transition and the lower temperature
kink related to the paramagnetic to antiferromagnetic phase transition
\cite{shan}. The structural phase transition temperature $T_{s}$
and the magnetic phase transition temperature $T_{m}$ are inferred
by the criterion shown in Fig. 1(c). The resistive anomaly in the
``parent'' compound is suppressed with Co doping and bulk SC shows
up at 10 K in the $x=0.025$ sample. $T_{c}$ then increases to 20
K in the $x=0.046$ sample and finally is suppressed
back to 16 K in the $x=0.065$ sample. Figure 1(d) shows the temperature
dependent susceptibility data taken at 5 Oe. Except for the $x=0.025$
sample, the ZFC $4\pi\chi$ data show a relatively sharp drop below
$T_{c}$ and saturate at low temperature with the magnitude of ZFC
$4\pi\chi$ at 2 K spreading from -110\% to -120\%, which are comparable
to the ones in the prototype Co doped Ba122 \cite{nico}, suggesting
100\% of SC shielding fraction here. Because of demagnetization
effects, when a thin disk of radius $a$ and thickness of $c\ll a$
is placed in a field $H\parallel a$, its intrinsic susceptibility
is given by $\chi_{\mathrm{intrinsic}}=\chi_{\mathrm{exp}}/(1-4\pi N\chi_{\mathrm{exp}})$,
where $N=0.5\pi c/a$ and $\chi_{\mathrm{exp}}$ is the value in Fig.
1(d). For the $x=0.025$ batch, the average $c/a$ ratio of the four
pieces measured is 0.10 and $4\pi\chi_{\mathrm{exp}}$ at 2 K is -90\%,
resulting in an average SC shielding fraction of 80\% at 2 K without
saturation. The small Meissner fraction inferred from the FC data
indicates strong flux pinning, a common feature in FBS. The representative
Hall coefficient data taken for the $x=0$ and $x=0.033$ samples
are presented in Fig. 1(e). Their negative values indicate that electron
carriers dominate the transport properties. Since $R_{H}=\frac{1}{nq}$,
where $n$ is the carrier density, the decrease of $|R_{H}|$ accompanied
with Co doping suggests larger $n$, consistent with the fact that
Co doping adds extra electron carriers into the system. Note that
the SC sample with the chemical formula Ca$_{0.82}$La$_{0.18}$FeAs$_{2}$
shows a positive Hall coefficient \cite{hall112}, indicating
that the SC in Ca$_{1-x}$La$_{x}$FeAs$_{2}$ arises from the hole
doping present in Ca$_{0.74}$La$_{0.26}$FeAs$_{2}$.

\begin{figure}
\centering \includegraphics[width=3.4in]{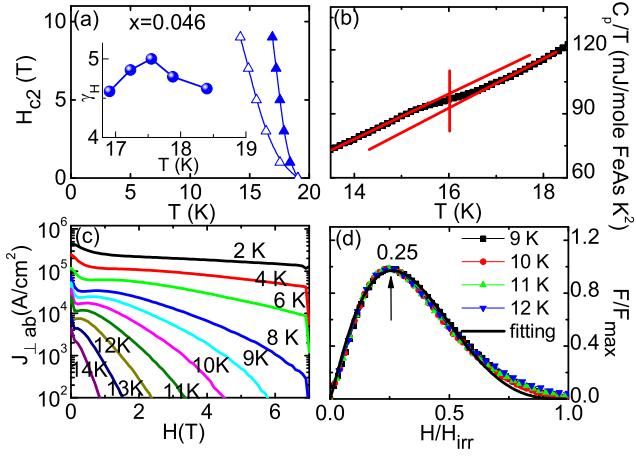} \protect\protect\caption{Ca$_{0.74(1)}$La$_{0.26(1)}$(Fe$_{0.954}$Co$_{0.046}$)As$_{2}$:
(a) The $H_{c2}$ data inferred using the 50\% criterion. Inset: the
anisotropy parameter of the upper critical field $\gamma_{H}=H_{c2}^{\perp ab}/H_{c2}^{\parallel ab}$
. (b) The $C_{p}/T$ vs. $T$. (c) The field dependent critical current
density at various temperatures. (d) Normalized pinning
force $f=F_{p}/F_{p,max}$ vs reduced field $h=H/H_{irr}$ at various
temperatures. The data were fitted by $f=Ah^{p}(1-h)^{q}$ with the
parameters $p=1.14$ and $q=3.24$.}

\label{fig:Fig2}
\end{figure}

As a representative, the superconducting properties of the $x=0.046$
sample are shown in Fig. 2. Figure 2(a) presents the upper critical
field $H_{c_{2}}(T)$, which is determined using the 50\% criterion
shown in Fig. 1(b). The anisotropy parameter of the upper critical
field $\gamma_{H}=H_{c2}^{\perp ab}/H_{c2}^{\parallel ab}$ is around
4.7. Since the effective mass anisotropy $\Gamma$ is
related to $\gamma_{H}$ by $\Gamma=\gamma_{H}^{2}=m^{*\perp ab}/m^{*\parallel ab}$,
the mass anisotropy is around 22. Note that $\gamma_{H}\approx4.7$
in 112 is smaller than $\gamma_{H}\approx8$ in the 10-3-8 compound,
which has similar FeAs interlayer distance but insulating spacer layers
\cite{ni1038,chen1038}. This suggests a stronger FeAs interlayer
coupling in 112 arising from the metallic spacer layers. Figure 2(b)
presents the temperature dependent specific heat $C_{p}$ taken at
H = 0 T. A bump in $C_{p}$ associated with the superconducting phase
transition appears, confirming bulk superconductivity in this sample.
By the equal entropy construction shown in Fig. 2(b), the heat capacity
jump $\Delta C_{p}/T|_{T_{c}}$ is 6.7 mJ/mole-Fe-$K^{2}$ at $T_{c}\approx$16
K. This value follows the Budko-Ni-Canfield(BNC) log-log plot quite
well \cite{BNC}, suggesting that $C_{p}\propto T_{c}^{3}$. Since
most of the FBS superconductors follow this BNC scaling \cite{sergy},
this may suggest an S$\pm$ pairing symmetry in CaLa112. The field
dependent current density $J_{c}$ at various temperatures is shown
in Fig. 2(c). We have calculated the critical current density based
on the Bean model \cite{Bean}, $J_{c}=20\frac{\triangle M}{w(1-\frac{w}{3l})}$,
where $\triangle$M = M$_{+}$ - M$_{-}$, and M$_{+}$ (M$_{-}$)
is the magnetization associated with increasing (decreasing) field;
$w$, $l$ is the width and length of the sample separately. At 2
K, $J_{c}$ reaches 2.2$\times$10$^{5}$ A/cm$^{2}$, comparable
to FeTe$_{0.5}$Se$_{0.5}$ and LiFeAs but lower than Ba$_{0}.6$K$_{0}.4$Fe$_{2}$As$_{2}$
and SmFeAsO$_{1-x}$ \cite{T. Taen}. To understand the mechanism
of the vortex motion, the normalized pinning force $f=F_{p}/F_{p,max}$
as a function of reduced field ($h=H/H_{irr}$) is plotted in Fig.
2(d), where $F_{p}=J_{c}\times\mu_{0}H$ \cite{dew}. All curves can
be scaled well and characterized by a maximum near $h$ = 0.25. This
value is smaller than most FBS, but close to $h_{max}$=0.28
for FeTe$_{0.7}$Se$_{0.3}$ \cite{bonura}, suggesting that both
surface pinning and small size normal core pinning contribute to the
vortex motion.

\begin{figure}
\centering \includegraphics[width=3.4in]{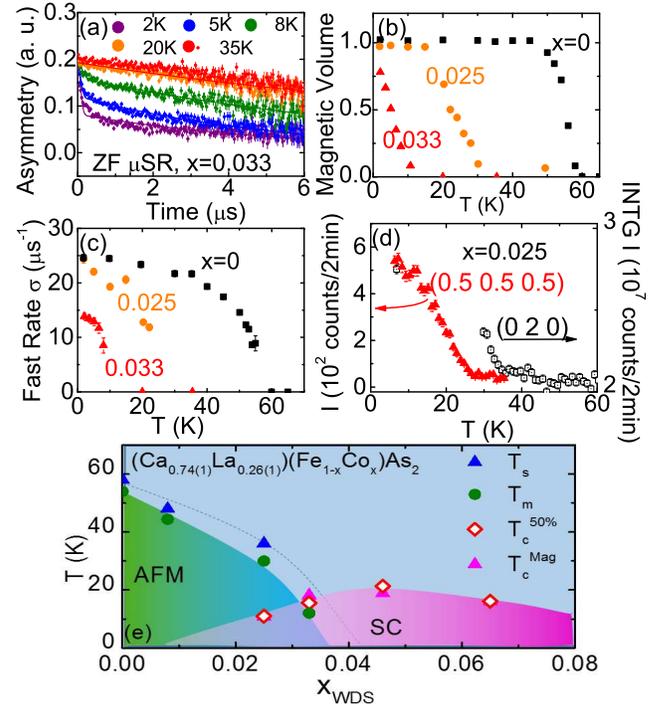} \protect\protect\caption{Ca$_{0.74(1)}$La$_{0.26(1)}$(Fe$_{1-x}$Co$_{x}$)As$_{2}$: (a)
The representative ZF $\mu$SR data of the $x=0.033$ sample. (b)
The temperature dependent ordered magnetic volume fraction $V_{\mathrm{mag}}$ determined
from the fitting of the ZF$\mu$SR asymmetry spectra. (c) The fast
transverse relaxation rate $\sigma$ inferred from the ZF $\mu$SR
asymmetry spectra. (d) The integrated intensity of the (0 2 0) nuclear
neutron peak and the intensity of the (0.5 0.5 0.5) magnetic neutron
peak. (e) The temperature-doping level ($T-x$) phase diagram of Ca$_{0.74(1)}$La$_{0.26(1)}$(Fe$_{1-x}$Co$_{x}$)As$_{2}$.
For $x=0,0.008$, $T_{s}$ and $T_{m}$ are the structural and magnetic
phase transition determined from the $dR/dT$, respectively. For $x=0.025$ and$ 0.033$,
$T_{m}$ is inferred by the ZF $\mu SR$ data. $T_{c}^{50\%}$ is the SC transition temperature
determined using the 50\% criterion shown in Fig. 1(b). $T_{c}^{Mag}$
is determined from the susceptibility data using the criterion shown
in Fig. 1(d).}

\label{fig:current}
\end{figure}

The interplay of structure, antiferromagnetism (AFM) and SC has been one of the most
studied topics in FBS, especially because spin fluctuations are believed
to play an important role in the pairing mechanism \cite{daice1111,ames122,dai122}.
To determine the phase diagram, and particularly to resolve if microscopic
coexistence between SC and AFM exists in the Co doped CaLa112, ZF
$\mu$SR measurements were performed to estimate the magnetic volume
fraction of the superconducting samples. As a representative, the
ZF asymmetry spectra of the $x=0.033$ sample at several temperatures
are shown in Fig. 3(a). Although no feature associated with a structural/magnetic
phase transition is observed in the resistivity measurements, upon
decreasing the temperature from 35 K, a fast relaxing front end is
absent at 20 K but clearly shows up at 8 K in Fig. 3(a), suggesting
the development of magnetic order in this sample at low temperature.
The ZF asymmetry spectra are fitted with the function
\begin{equation}
A_{ZF}(t)=A[f_{T}\exp{(-\frac{1}{2}(\sigma t)^{2})}+(1-f_{T})\exp{(-\lambda t)}].
\end{equation}
Here $f_{T}$ is the transverse function and denotes the fast relaxing
component coming from the static magnetic order. $\sigma$ is the
fast transverse relaxation rate of $\mu^{+}$. $(1-f_{t})$ is the longitudinal fraction, representing the fraction of
muons with spins parallel to the local magnetic field in the
region with static magnetic order and/or muons in the paramagnetic volume.
The longitudinal fraction only undergoes the spin lattice relaxation
with the rate $\lambda = 1/T_{1}$.

The Gaussian relaxation form arises from the magnetic field produced
by randomly oriented magnetic moments, as expected in our Co-CaLa112
sample. For an ideally isotropic and fully ordered system, $f_{T}$
is expected to be 0.67 while the $f_{T}$ values at 2 K in $x=0$,
0.025 and 0.033 samples are 0.75, 0.72 and 0.58, respectively. The
slightly larger values in the $x=0$ and 0.025 samples likely arise
from the field anisotropy in these randomly packed plate-like single
crystals. By setting the magnetic volume in the $x=0$ sample to be
100\% as proved reasonable in Ref. \cite{shan}, we calculate the
magnetic volume of the other samples, as shown in Fig. 3(b). For the
$x=0.025$ sample, due to the sample inhomogeneity, around 6.4\% of
the sample is already magnetic at 50 K, but for the rest of the sample,
AFM develops below 31 K and the magnetic volume
fraction $V_{\mathrm{mag}}$ saturates below 20 K with $V_{\mathrm{mag}}\approx$96\%
at 2 K. For the $x=0.033$ sample, AFM appears below 12 K and
$V_{\mathrm{mag}}$ increases up to 76\% at 2 K without
saturation. The fast transverse relaxation rate $\sigma$ is shown in Fig.
3(c). Since $\sigma$ is proportional to the size of the magnetic
moment, which is 1.0 $\mu_{B}$/Fe in the $x=0$ sample
as revealed by neutron diffraction \cite{shan}, assuming that the
$\mu^{+}$ sites remain similar in these measured samples, the magnetic
moment in the $x=0.033$ sample is 0.6$\mu_{B}$/Fe while it remains
1.0$\mu_{B}$/Fe in the $x=0.025$ sample. The neutron diffraction
data taken on a $x=0.025$ single crystal are presented in Fig. 3(d).
The integrated (0 2 0) nuclear peak intensity sharply increases below
35 K, similar to the one in the $x=0$ sample \cite{shan}, marking
the occurrence of a structural phase transition. The intensity of
the (0.5 0.5 0.5) magnetic peak abruptly increases below 28 K, indicating
the development of long range magnetic ordering, which is consistent
with the $\mu$SR data shown in Fig. 3(b).

Based on the data we discussed above, a $T-x$ phase diagram is constructed
in Fig. 3(e). Upon Co doping, the structural/magnetic phase transitions
are suppressed and SC up to 20 K emerges. For the $x=0.025$ sample,
the magnetic volume fraction determined from the $\mu$SR
data is 96\%. The SC volume fraction determined
from the susceptibility data is 80\%. For the $x=0.033$ sample, the
SC volume fraction is 100\% while the magnetic volume
fraction is at least 76\% at 2 K, which has the trend
to sharply increase further at even lower temperature. Therefore,
there is clearly microscopic coexistence between SC and AFM
in $x=0.025$ and $0.033$ samples, similar to most doped Ba122
compounds and the Ca$_{1-x}$La$_{x}$FeAs$_{2}$ \cite{coexist1,coexist2,coexist3,coexist4,coexist5,coexist6,coexist7,nmr112}.

\begin{figure}
\centering \includegraphics[width=3in]{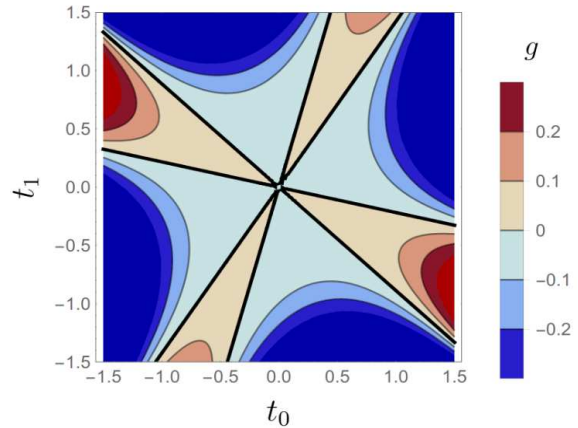} \protect\protect\caption{The coefficient $g$ as function of the tight-binding
parameters $t_{0}$ and $t_{1}$ in Eq. (\ref{Fig:TransitionInTSpace}),
under the constraint $t_{2}=-t_{0}-t_{1}$ to maintain the perfect
nesting condition at $k_{z}=0$. Note that $g=0$ at the black lines,
whereas $g<0$ in the blue-shaded region (implying AFM-SC microscopic
phase coexistence) and $g>0$ and in the red-shaded region (implying
AFM-SC macroscopic phase separation).}

\label{Fig:TransitionInTSpace}
\end{figure}

It is instructive to compare these results for the 112 Ca$_{1-x}$La$_{x}$FeAs$_{2}$
compound with those for the 10-3-8 Ca$_{10}$(Pt$_{3}$As$_{8}$)((Fe$_{1-x}$Pt$_x$)$_{2}$As$_{2}$)$_{5}$
system. Both of them have similar atomic constituents and very close
values for the interlayer FeAs distance. However, in contrast to the
112, where an extended AFM-SC coexistence region emerges, the 10-3-8
shows no coexistence -- or at best a very limited region of coexistence
\cite{La1038,ruslan1038}. One of the most salient differences between
these two classes of compounds is the fact that in the 10-3-8, the
spacer layer is insulating, whereas in the 112 it is metallic. This
difference is manifest, for instance, in the larger $H_{c2}$ anisotropy
of the former ($\gamma_{H}\approx8$) over the latter ($\gamma_{H}\approx4.7$).
Presumably, the existence of a metallic spacer layer enhances the FeAs interlayer
coupling in the 112, making it more three-dimensional. Although several
factors could be at play, it is tempting to attribute the presence
of the AFM-SC coexistence region in these systems to the difference
in their degree of three-dimensionality promoted by the distinct characters
of the spacer layers.

In order to investigate whether the more pronounced
$k_{z}$ band dispersion of the 112 material favors the microscopic
coexistence of AFM and SC, we consider a toy two-band model widely
employed in the FBS to study the competition of AFM and SC \cite{vvc,parker09,fs}.
This model is characterized by a hole-pocket with dispersion $\xi_{h}$
at the center of the Brillouin zone and an electron-pocket with dispersion
$\xi_{e}$ at the corner of the Brillouin zone. The fate of the competing
AFM-SC phases is determined by a single coefficient $g$, which depends
on the quartic coefficient of the microscopically-derived Ginzburg-Landau
expansion \cite{supp}: if $g>0$, the competition between the phase
is so strong that there is no coexistence, whereas if $g<0$, their
competition is weak enough to allow them to coexist microscopically.
In the hypothetical perfect nesting limit, $\xi_{e}=-\xi_{h}$, $s^{+-}$
SC and AFM are at the verge of coexistence or macroscopic phase separation,
with $g=0$. Deviations from perfect nesting then determine whether
$g$ becomes positive or negative. Previously, deviations of perfect
nesting arising from the in-plane band dispersions were studied \cite{fs,vvc}.
Here, we consider deviations arising from the out-of-plane band dispersion,
and write $\xi_{e}=-\xi_{h}+\lambda b\left(k_{z}\right)$, with the
general $k_{z}$-dispersion:
\begin{equation}
b\left(k_{z}\right)=t_{0}+t_{1}\cos(p_{z})+t_{2}\cos(2p_{z})
\end{equation}

To ensure that the system is nested at $k_{z}=0$,
the tight-binding coefficients are constrained to $t_{2}=-t_{0}-t_{1}$.
In Fig. 4 \cite{supp}, we compute the value of $g$ in the $\left(t_{0},t_{1}\right)$
parameter space. Clearly, $g<0$ in a wide region of the parameter
space, showing that in general the $k_{z}$ dispersion can promote
microscopic coexistence between SC and AFM. Thus, this simple
calculation lends support to the idea that the enhanced interlayer
coupling in the 112 promoted by the metallic spacer contributes to
the stabilization of a regime of microscopic AFM-SC coexistence in
the phase diagram.

In summary, we have mapped out the $T-x$ phase diagram of the Ca$_{0.74(1)}$La$_{0.26(1)}$(Fe$_{1-x}$Co$_{x}$)As$_{2}$
superconductors. Microscopic coexistence between AFM and SC exists
in this FBS. A phenomenological two-band model suggests the dispersion
along $k_{z}$ direction may favors AFM-SC microscopic coexistence
over phase separation.

\textit{Acknowledgments.}Work at UCLA was supported by the NSF DMREF
DMR-1435672. Work at ORNL$^{\prime}$s High Flux Isotope Reactor was
sponsored by the Scientific User Facilities Division, BES, DOE. Work
at Columbia and TRIUMF was supported by the NSF DMREF DMR-1436095,
PIRE project IIA 0968226, DMR-1105961 and JAEA REIMEI. M.S. acknowledges
the support from the Humboldt Foundation. R.M.F. is supported by the
U.S. Department of Energy, Office of Science, Basic Energy Sciences,
under Award No. DE-SC0012336.

\end{document}